%% file: main.tex
\documentclass[10pt, conference]{ieeetran}

\usepackage{amsmath,amssymb,amsfonts}
\usepackage{balance}
\usepackage{booktabs}
\usepackage{caption}
\usepackage{comment}
\usepackage{dirtytalk}
\usepackage{graphicx}
\usepackage{hyperref}
\usepackage{cleveref}
\usepackage{subcaption}
\usepackage{textcomp}
\usepackage{url}
\usepackage{xcolor}

\usepackage{cite}
\usepackage{lipsum}  
\usepackage{float}
\usepackage{array}
\usepackage{adjustbox}
\usepackage{pbox}
\usepackage{hyperref}
\usepackage{blindtext}
\usepackage{multirow}

\graphicspath{{\subfix{../pictures/}}}

\author{
 \IEEEauthorblockN{Conrado Boeira$^\star$, Antor Hasan$^\star$, Khaleda Papry$^\star$, Yue Ju$^\diamond$, Zhongwen Zhu$^\diamond$, Israat Haque$^\star$}
 
\IEEEauthorblockA{$^\diamond$Ericsson GAIA, Montreal, Canada}
\IEEEauthorblockA{$^\star$Department of Computer Science, Dalhousie University, Canada}
}

\begin{document}


\title{ A Calibrated and Automated Simulator for Innovations in 5G }

\maketitle

\begin{abstract}
The rise of 5G deployments has created the environment for many emerging technologies to flourish. Self-driving vehicles, Augmented and Virtual Reality, and remote operations are examples of applications that leverage 5G networks' support for extremely low latency, high bandwidth, and increased throughput. However, the complex architecture of 5G hinders innovation due to the lack of accessibility to testbeds or realistic simulators with adequate 5G functionalities. Also, configuring and managing simulators are complex and time consuming. Finally, the lack of adequate representative data hinders the data-driven designs in 5G campaigns. Thus, we calibrated a system-level open-source simulator, Simu5G, following 3GPP guidelines to enable faster innovation in the 5G domain. Furthermore, we developed an API for automatic simulator configuration without knowing the underlying architectural details. Finally, we demonstrate the usage of the calibrated and automated simulator by developing an ML-based anomaly detection in a 5G Radio Access Network (RAN).


\end{abstract}

\input{sections/introduction}

\input{sections/background_and_motivation}

\input{sections/related_work}

\input{sections/methodology}

\input{sections/automation}
\input{sections/use-cases}

\input{sections/conclusion}

\bibliographystyle{IEEEtran}
\balance
\bibliography{refs}

\end{document}

%% file: sections/introduction.tex
\section{Introduction}

The emergence of various IoT applications (e.g., AR/VR, autonomous vehicles) and their demand motivated both industry and academia to revisit the cellular network design. Fifth-generation (5G) cellular networks evolved to support these emerging applications with diverse latency and bandwidth needs. 5G New Radio (NR) targets to achieve connectivity with low latency and high speed (Gigabits) by employing both sub-GHz and mmWave frequency bands \cite{panwar2016survey}. In particular, Mobile Broadband (eMBB) offers enhanced bandwidth for applications like AR/VR, whereas autonomous vehicles may rely on ultra-Reliable Low Latency (uRLLC). Finally, massive Machine Type Communication (mMTC) supports low-power massive machine-to-machine communications like industry 4.0. eMBB also supports regular users to provide high bandwidth and better Internet access quality.  

5G deploys functionalities like network slicing, support for millimeter wave frequency band, and different coding schemes, which make it highly efficient at the cost of management and control complexities \cite{budhdev2021fsa,jain2022l25gc,zhao2022seed,chukhno2022placement,sukhmani2018edge}. Thus, the operators need to measure and test new functionalities with appropriate parameter tuning before being deployed in the production line. Simulators enable faster assessment and testing of new 5G capabilities \cite{gkonis2020comprehensive}. Furthermore, researchers can evaluate their ideas without needing to access real deployments.  

However, these simulators must be realistic to offer a dependable evaluation platform. Thus, the 3rd Generation Partnership Project (3GPP), the primary standards organization for mobile telecommunications, defined the parameters to be used and their calibrating and tuning guidelines to realize realistic simulators.

There are open source 5G simulators (e.g., 5G-Lena\cite{patriciello2019e2e}, Vienna Simulator \cite{muller2018flexible}, and  5G-air-simulator \cite{martiradonna20205g}) for rapid prototype implementation and evaluation. One way to make these simulators dependable is conducting \textit{calibration} following 3GPP specifications. ITU-R \cite{ituimt} has defined the technical requirements and measurements for radio interface technologies following 3GPP-defined reference scenarios and guidelines. Leading telco companies (e.g., Ericsson, Huawei, and Nokia) calibrated their simulators with this standard. Thus, we can calibrate a new simulator and compare the results against these existing ones to verify the calibration validity. For example, Wise and 5G-Lena are calibrated simulators following 3GPP specifications \cite{koutlia2022calibration,jao2018wise}. However, the former is proprietary, and the latter does not support functionalities like handover, which limits the scope of these simulators. 

Furthermore, the complexity and scale of 5G networks initiated data-driven systems designs. However, many providers may not label or offer public access to their data \cite{chen2020active}, which creates a barrier in designing data-driven solutions in 5G. The calibrated simulators with essential data collection interfaces can alleviate data accessibility issues and catalyze innovations. However, another barrier to that innovation is the steep learning curve while using simulators like 5G-Lena and Vienna.   


This paper fills the above three identified gaps with three contributions. First, we calibrate an end-to-end system-level simulator Simu5G \cite{nardini2020simu5g} with various 5G deployment capabilities, e.g., support for user handover, dual-connectivity, and D2D communications following 3GPP guidelines. Specifically, we consider both the urban and rural deployment for eMBB, which aims to offer end users better throughput and bandwidth, with a mid-band configuration defined by 3GPP. Next, we automate the calibrated simulator and propose a YAML-based API for configuration specification. Thus, users need only to provide the topological information to get the necessary simulator configuration. Finally, we demonstrate the usability of the calibrated simulator for data-driven application assessment. We develop a neural network-based anomaly detection model and assess its performance on the generated faulty and normal data samples (collection of Key Performance Indicators (KPIs)) in an urban deployment. To the best of our knowledge, this is the first holistic approach of a 5G simulator calibration and automation to speed up innovations. The contributions of the paper are summarized as follows.

\begin{itemize}
    \item \textbf{Simu5G calibration}: we calibrate this simulator following 3GPP specifications and guidelines for calibration in urban and rural deployments to ensure dependability. 
    
    \item \textbf{Simulation automation}: we propose a YAML-based tool to automatically generate simulation configuration and tasks with the calibrated parameters for rapid prototyping.
    
    \item \textbf{ML use case}: we demonstrate an ML-based anomaly detection use case to showcase the calibrated simulator's usage in assessing data-driven designs. 

    \item \textbf{Open source code}: we share the developed solutions over Git Repo \cite{Boeira_Calibration-and-Automation-of-a-5G-Simulator_2024} for reproducibility and extension. 
    
\end{itemize}

%% file: sections/background_and_motivation.tex
\section{Background}

This section presents the necessary background to understand the proposed work.

\textbf{KPIs for calibration.}
3GPP has defined two KPIs for simulator calibration: \textit{Downlink Coupling Gain} and \textit{Downlink Geometry}. The former can be defined as the difference between received and sent signal strength and takes into consideration the path loss, antenna gains, fast fading, and analog beamforming gains when applicable, i.e., $Coupling Gain [dB] = P^T - P^R$ \cite{series2017guidelines}. The downlink geometry is the ratio between the signal power received and the sum of noise and interference experienced during propagation. It is, therefore, a measurement of Wideband SINR and can be defined as $SINR = \frac {P^R}{I+N}$, where $N$ is the noise power and $I$ is the sum of interference powers.

\textbf{Common faults in RAN.} We consider the 3 most common faults: too-late handover, excessive power reduction, and interference.

\textit{Too-late handover}: can be defined as an unsuccessful Handover (HO) process. HO is a common function in mobile networks aiming to provide mobility for users, which allows a user to change a connection from one base station to another to maintain consistent and good signal quality while moving across base stations. 
In the HO process, users constantly send the signal strength from neighboring base stations to the associated gNB, and it checks if the conditions for a handover event are met. These Handover Events are conditions that include multiple parameters such as specific thresholds and offsets. If these conditions are met for a time-to-trigger duration, the handover process is started. The most widely used event is called A3, where the neighboring cell’s signal becomes stronger than the serving cell’s signal by a specific margin\cite{lee2020prediction}. Base stations maintain two main configuration parameters: \textit{hysteresis}, the margin between signal strengths, and \textit{time-to-trigger}. When the difference of a specific KPI, commonly RSRP, between the serving and neighbor base stations exceeds \textit{hysteresis} value for a period longer than the \textit{time-to-trigger}, the HO process is triggered by the serving gNB. 
If a base station is incorrectly configured, it may suffer from too-late handover, where the HO process is not triggered on time and users can experience worse signal quality. 

\textit{Excessive power reduction}: happens when the power produced by a cell is reduced to a level where it is not able to provide a planned performance requirement. This issue could occur due to poor configuration settings or wiring problems \cite{gomez2015methodology}.

\textit{Inter-Cell Interference}: is one significant cause of issues in telecommunication networks, and it happens when users of neighbor cells use the same frequency. Thus, the intended receiving signals are mixed with the interfering signals and distort the reception \cite{siddiqui2021interference}. Such interference can occur if users from the neighboring cells use the same frequency or if those cells are incorrectly configured. 






%% file: sections/related_work.tex
\section{Related Work}
\label{sec:related-work}


\textbf{Simulator calibration and automation.} 
There are industrial proprietary and open-source 5G simulators. Proprietary simulators are generally used by industries and not available for public usage, e.g., NOMOR’s Simulator \cite{yu2018imt}. Hence, our work focuses on open-source 5G simulators due to accessibility and reproducibility. Open-source simulators are link-level or end-to-end system-level, where the former simulates a point-to-point communication link focusing on physical layer functionalities \cite{sun2017novel, dominguez2016gtec}. System-level ones operate and interact across multiple protocol layers and model multi-node networks with different deployment configurations. Specifically, these simulators can handle end-to-end discrete events like packet arrivals/departures from nodes. 

We focus on the open-source 5G system-level simulators and list them in Table~\ref{table1}. Among these simulators, 5G-LENA \cite{koutlia2022calibration},  WiSE \cite{jao2018wise}, and 5G Toolbox \cite{xue20225g} are fully calibrated following 3GPP guidelines along with the calibration outcomes. On the other hand, 5G-Air-simulator \cite{martiradonna20205g} and PycheSim \cite{pereyra2021py5chesim} compare their calibrated outcome with existing calibrated models instead of comparing with 3GPP.   For example, 5G-Air is compared with industry-level simulators (e.g., Nokia, Huawei, Ericsson, Samsung) without comparing with 3GPP. Finally,  SyntheticNET \cite{zaidi2020syntheticnet} does not have any calibration results.  

Among the above simulators, only 5G-LENA supports full protocol stack and end-to-end simulation with rural-eMBB and dense urban-eMBB calibration environments. However, it fails to provide D2D communication and dual connectivity support and mainly lacks handling handovers. 5G-air simulator, WiSE, and SyntheticNET are capable of handling handover but do not support D2D communication, dual connectivity,  or full-stack protocol. NS-3 mmWave and 5G K\-SimNet support dual connectivity and handovers, where the former is not calibrated following 3GPP and the latter lacks the full protocol stack. Finally, SyntheticNet included some AI-based automation like importing BS\/UE specific configuration information from a database. 
To fill the above-mentioned gaps and provide user-friendly network configuration automation with calibrated parameters, we first calibrate Simu5G following 3GPP guidelines and then introduce an automated configuration tool to realize a realistic 5G NR.

\begin{table*}[t]
    \centering
    \begin{tabular}{|p{3.2cm}|p{2cm}|p{1.3cm}|p{1.5cm}|p{1.5cm}|p{1.8cm}| p{1.7cm}|p{1.3cm}|} \hline
    
\textbf{Simulator
}&  \textbf{D2D \newline comm.} &  \textbf{Dual  
  \newline connectivity} &  \textbf{End-to-end \newline simulation} &  \textbf{Handover \newline support} &\textbf{Full \newline protocol stack} &  \textbf{Calibrated \newline with 3GPP}&  \textbf{Automated}  \\ 
\hline
5G-Lena \cite{koutlia2022calibration} & $\times$   & $\times$ & \checkmark & $\times$ & \checkmark & \checkmark & $\times$ \\
\hline
ns-3 mmWave \cite{mezzavilla2018end} & no mention   & \checkmark  & \checkmark & \checkmark & \checkmark & $\times$ & $\times$ \\
\hline
5G -air-simulator \cite{martiradonna20205g}& 
$\times$  & $\times$ & \checkmark & \checkmark & $\times$ & no comparison & $\times$\\
\hline
WiSE \cite{jao2018wise} & 
$\times$  & $\times$ & no mention & \checkmark & $\times$ & \checkmark & $\times$\\
\hline
5G Vienna SL \cite{muller2018flexible} & 
 \checkmark  & $\times$ &  $\times$ & $\times$ & $\times$ &  $\times$ & $\times$\\
\hline
5G Toolbox by Matlab \cite{xue20225g} & 
 \checkmark  & no mention &   \checkmark & $\times$ & $\times$ &  \checkmark & $\times$\\

\hline
5G K-SimNet \cite{chen2020root} & 
no mention & \checkmark  &   \checkmark & \checkmark  & $\times$ &  $\times$ & $\times$\\
\hline
Py5cheSim \cite{pereyra2021py5chesim} & 
 $\times$ & $\times$ &   \checkmark & $\times$ & $\times$ &  no comparison  & $\times$\\
\hline
SyntheticNET \cite{zaidi2020syntheticnet} & 
\checkmark  & no mention &   $\times$ & \checkmark  & $\times$ &  no results & partial\\
\hline
Simu5G\_calibrated & \checkmark  &  \checkmark &   \checkmark &  \checkmark &  \checkmark &  \checkmark &  \checkmark \\
\hline
\end{tabular}
\caption{ 5G open-source system-level simulators. }
\label{table1}
\end{table*}

\textbf{Anomaly detection and Root-Cause Analysis (RCA).}
Meng Chen \textit{et al.} proposed a pool-based active learning to label data and then apply XGBOOST for fault diagnosis in LTE systems without considering any spatio-temporal data relationship  \cite{chen2020root}. Zhang \textit{et al.} filled that gap and combined Convolutional Neural Network (CNN) and Long Short-Term Memory (LSTM) to handle the spatio-temporal dependencies of fault prediction problems in LTE RAN networks \cite{zhang2019self}. An ensemble-based LSTM model for multiple faults in 5G RAN is developed in \cite{sundqvist2020boosted}. The model's performance is evaluated using real traffic generated in a 5G testbed without providing public access. Multilayer Perceptron, Decision Tree, and Support Vector Machine are used to detect anomaly in 5G Open RAN (O-RAN) in \cite{alves2023machine}. A procedural-based anomaly detection model in 5G RAN is proposed in \cite{sundqvist2023robust} to reduce the detection time without degrading the performance. Finally, a Graph Neural Network (GNN)-based fault detection system is proposed in \cite{yen2022graph}.

A key demand of the above data-driven designs is to have access to adequate (e.g., large volume for deep neural networks) data on time. However, the above works falls short in offering enough open-access data or require expert support to directly interact with the network and label faults, making it infeasible for many researchers. This work fills the gap by introducing a calibrated and automated 5G simulator to generate the required amount of data in a reasonable time without spending too much time on understanding and configuring the simulator.

%% file: sections/methodology.tex
\section{Calibration}
\label{sec:calibration}



ITU defines \cite{ituimt} the process of calibration of different radio technologies. They proposed multiple standard deployment scenarios. Also, 3GPP defined reference results based on various industrial simulator calibrations. There are three different deployment scenarios, eMBB, mMTC, and uRLLC, that form five different environments: Indoor Hotspot-eMBB, Dense Urban-eMBB, Rural-eMBB, Urban Macro–mMTC, and Urban Macro–URLLC. This work considers eMBB scenarios as they are among the most common to offer mobile broadband to end users. Specifically, we consider dense urban-eMBB and rural-eMBB scenarios, where the urban scenario can have evaluation configurations A and B.

In contrast, Rural one can use configurations A, B, and C. These configurations differ in carrier frequency, bandwidth, and other parameters. For instance, in the case of the Urban scenario, configurations A and B use a 4 GHz mid-band and a 30 GHz high-band, respectively. On the contrary, in the Rural scenario, both A and C use a low-band 700 MHz, and B is a 4 GHz mid-band. We chose the most widely used mid-band configurations: A for urban and B for rural deployment.   



\subsection{Methodology}

We need to calibrate two KPIs Coupling Gain and Wideband SINR as part of the calibration process following 3GPP guidelines. Thus, we collect these two KPI data for the chosen deployment scenarios and configurations to compare against the 3GPP provided reference data. In particular, we compare the cumulative distribution function (CDF) of the collected data to the 3GPP and other industry provided reference CDFs for the above two KPIs. In addition, we use the Kolomogorov-Smirnov test (KS test) \cite{massey1951kolmogorov} to measure how close our calibrated and 3GPP distributions are, where KS value close to 0 indicates a better match. Also, we compare this value with the other industrial submissions.

However, the standard configurations of the simulator do not match the standards by default, as there are a myriad of parameters that need to be perfectly set so that the results are aligned with the 3GPP provided values. Thus, in order to achieve a good calibration, we must set and fine-tune all required parameters following 3GPP guidelines. Also, we may require adding 5G functionalities in the chosen simulator to realize the required environments. The key challenges toward that goal include having in-depth knowledge about the 5G technologies and expertise on the simulator.



In the calibration process, we first set the parameters that 3GPP defined (as presented in Table~\ref{tab:scenario}). In addition, we must tune some parameters following 3GPP guidelines and the comments from the partner organization. In our considered scenarios, these tuning parameters include the maximum deployment distance between the user and the base station, the building height, the thermal noise, and the number of background users. These parameters impact the chosen two KPIs but have no defined values indicated in the 3GPP guideline. For example, the minimum deployment distance from users to base stations is listed in 3GPP as 10m; however, there is no clear indication of how far these users can be positioned from base stations. Thus, we first follow the literature, industrial recommendations \cite{narayanan2022comparative, koutlia2022calibration,3gpp.138.133}, and our partner's guideline to identify a range of values for the above candidate parameters. Then, we tune one candidate parameter at a time while keeping the rest unchanged. The process is repeated until all parameters are tuned, presented in Table \ref{tab:scenario}.

\subsection{Implementation}

\begin{figure}[H]
    \centering
    \includegraphics[width=0.18\textwidth]{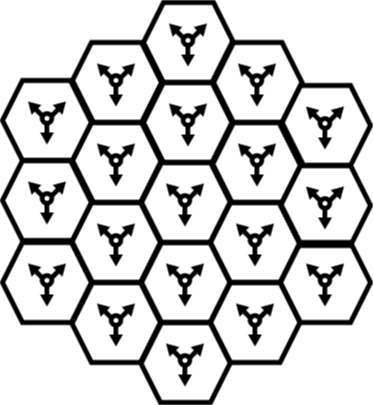}
    \caption{An example cellular network deployment with hexagonal grid and tri-sector antennas.}
    \label{fig:hexagonal_grid}
\end{figure}

We must consider a standard topology for the calibration process, which usually consists of \textit{tri-sector base stations} distributed over a hexagonal grid. Each site consists of 3 base stations and the area is divided into 19 hexagons to have a total of 57 cells (Fig.~\ref{fig:hexagonal_grid}). The list of parameters used in the chosen scenarios is presented in Table \ref{tab:scenario}, where the height of indoor users is defined by the following equation:
\begin{equation}
h_{UT} = 3(nfl -1) + 1.5
\end{equation}
where \(nfl \smash{\sim} uniform(1, Nfl)\)
  and $Nfl \sim uniform(4, 8)$ \cite{Huawei}.


\begin{table*}[t]
    \centering
    \begin{tabular}{|p{4cm}p{5cm}p{5cm}p{2cm}|}
        \hline
        Parameter & Value for Urban Scenario & Value for Rural Scenario & Reference \\
        \hline
        Carrier frequency & 4 GHz & 4GHz & 3GPP \\
        Bandwidth & 10 MHz & 10 MHz & 3GPP\\
        Inter-site distance & 200 m & 1732m & 3GPP\\
        Sectors & 30/150/270 degrees & 30/150/270 degrees & 3GPP\\
        BS antenna height & 25m & 35m &3GPP\\
        UE antenna height & outdoor UEs: 1.5 m, indoor UEs: Equation 1 & 1.5 m & 3GPP\\
        BS transmit power & 41 dBm/sector & 46 dBm/sector & 3GPP \\
        UE transmit power & 23 dBm & 23 dBm & 3GPP\\
        UEs deployment & 10 UEs per sector, randomly dropped &10 UEs per sector, randomly dropped & 3GPP\\
        Propagation model & 3GPP RMa TR 38.901 & 3GPP RMa TR 38.901 & 3GPP\\
        BS antenna element gain & 8 dBi & 8 dBi & 3GPP\\
        UE antenna element gain & 0 dBi & 0 dBi & 3GPP\\
        BS noise figure & 5 dB & 5 dB & 3GPP\\
        UE noise figure & 7 dB & 7 dB & 3GPP\\
        Device deployment & 80\% indoor, 20\% outdoor & 50\% indoor, 50\% outdoor & 3GPP\\
        High/low penetration loss building type  & 80\% low loss–20\% high loss & 100\% low loss & 3GPP\\
        UE Mobility Model & Random Waypoint Model & Random Waypoint Model & 3GPP \\ 
        UE Speed & indoor UEs: 3 km/h, outdoor UEs: 30 km/h  & indoor UEs: 3 km/h, outdoor UEs: 120 km/h  & 3GPP\\
        Traffic Load & Full Buffer & Full Buffer & 3GPP \\
        Maximum BS-UE deployment distance & 50 m & 200m & Calibrated\\
        Building Height & 22.5 m & 10 m & Calibrated\\
        Number of Cell Sites collecting data & 7 & 7 & 5G-Lena\cite{koutlia2022calibration} \\
        Number of background users & 10 per Background Cell & 10 per Background Cell & Calibrated \\
        Thermal Noise & -81 dBM & -82 dBM & Calibrated \\
        \hline
    \end{tabular}
    \caption{Parameters used in the Urban and Rural scenarios.}
    \label{tab:scenario}
\end{table*}

Furthermore, we need to revisit the following functions as part of the calibration process to align with the 3GPP standard. 

\textit{High penetration loss}: 3GPP defines two types of loss that may occur while signals propagate through walls and/or obstacles \cite{3gpp.138.901}. These losses are high and low penetration loss mainly depends on the material of obstacles. 3GPP recommends associating some users with high loss, whereas the rest with low loss. Specifically, Equation~\ref{low} and Equation~\ref{high} define the low and high loss, respectively \cite{3gpp.138.901}. However, in Simu5G, there was an error in the material coefficient used for each equation, where the IIR glass and standard glass coefficient were mixed, leading to inaccurate penetration losses. Thus, we fix the issues as part of the calibration process.

\begin{equation} \label{low}
PL = 5 - 10 log_{10}(0.3*10^{\frac{-L_{glass}}{10}} +  0.7*10^{\frac{-L_{concrete}}{10}}) 
\end{equation}

\begin{equation} \label{high}
PL = 5 - 10 log_{10}(0.7*10^{\frac{-L_{IIRglass}}{10}} +  0.3*10^{\frac{-L_{concrete}}{10}}) 
\end{equation}




\textit{Band usage}: The band usage implementation in Simu5G had an issue in the initialization phase. Specifically, the simulator used to show the usage status as \textit{true} right after the initialization despite not having started using it. Thus, we added a status of band usage as being used or not after the initialization phase. This change also helped improve the interference as we could accurately use available bands.

\subsection{Evaluation Results}

\subsubsection{Urban Deployment}
\begin{figure*}[!ht]
\centering
  \includegraphics[width=0.88\linewidth]{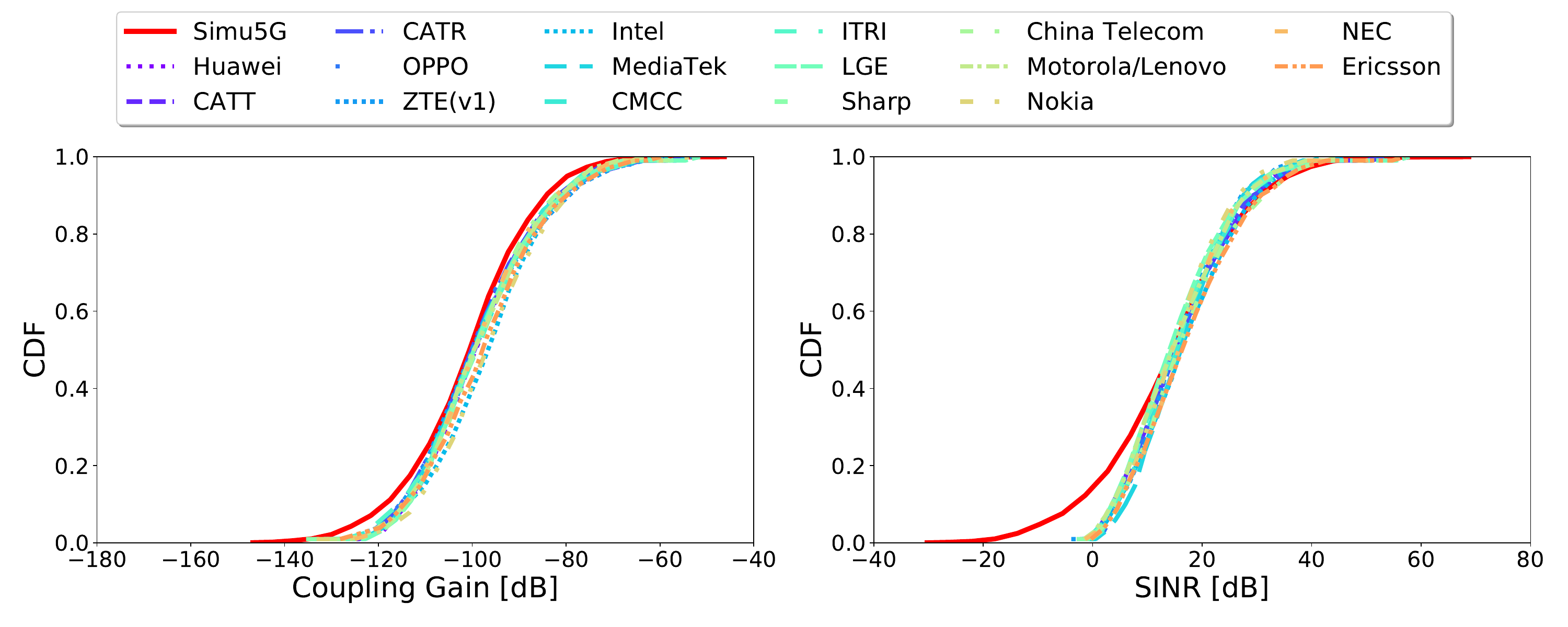}
\caption{Simu5G results compared to 3GPP submissions with max UE distance of 50m for the Urban eMBB.}
\label{fig:urban}
\end{figure*}

The CDF of Coupling Gain and Wideband SINR in Simu5G is compared against 3GPP and industrial submissions in Fig.~\ref{fig:urban}. In the case of Coupling Gain, we see a close alignment of calibrated Simu5G with the reference submissions with a marginal discrepancy. 
In the case of SINR, the average discrepancy between Simu5G and 3GPP is below 2 dB, which is within the acceptable range \cite{Huawei}. However, we observe a discrepancy in the lower range of SINR, which could be because Simu5G does not have a precise beamforming mechanism when calculating signal propagation. As precise beamforming can reduce interference by eliminating inter-device interference and allow for better frequency re-utilization \cite{rao20215g}, the lack of it can lead to low SINR values.

Furthermore, we present the KS test results in Table \ref{tab:ks_urban}. Specifically, we measure and compare the Simu5G KS scores to the maximum and average values of 3GPP submissions. The results show a similar trend to that of the CDF. The Coupling Gain score is between the maximum and average. The SINR score is slightly lower than the maximum but slightly higher than the average due to the higher quantity of lower-end values.


\begin{table}[!ht]
    \centering
    \resizebox{\columnwidth}{!}{%
    \begin{tabular}{|c|c|c|}
        \hline
        Name & Coupling Gain Score & Wideband SINR Score   \\
        \hline
        Simu5G & 0.065 & 0.138 \\
        Average Value in 3GPP & 0.031 & 0.070 \\
        Maximum Value in 3GPP & 0.089 & 0.287 \\
        \hline
    \end{tabular}}
    \caption{KS test results in the Urban eMBB.}
    \label{tab:ks_urban}
\end{table}

\subsubsection{Rural Deployment}
\begin{figure*}[!ht]
\centering
  \includegraphics[width=0.88\linewidth]{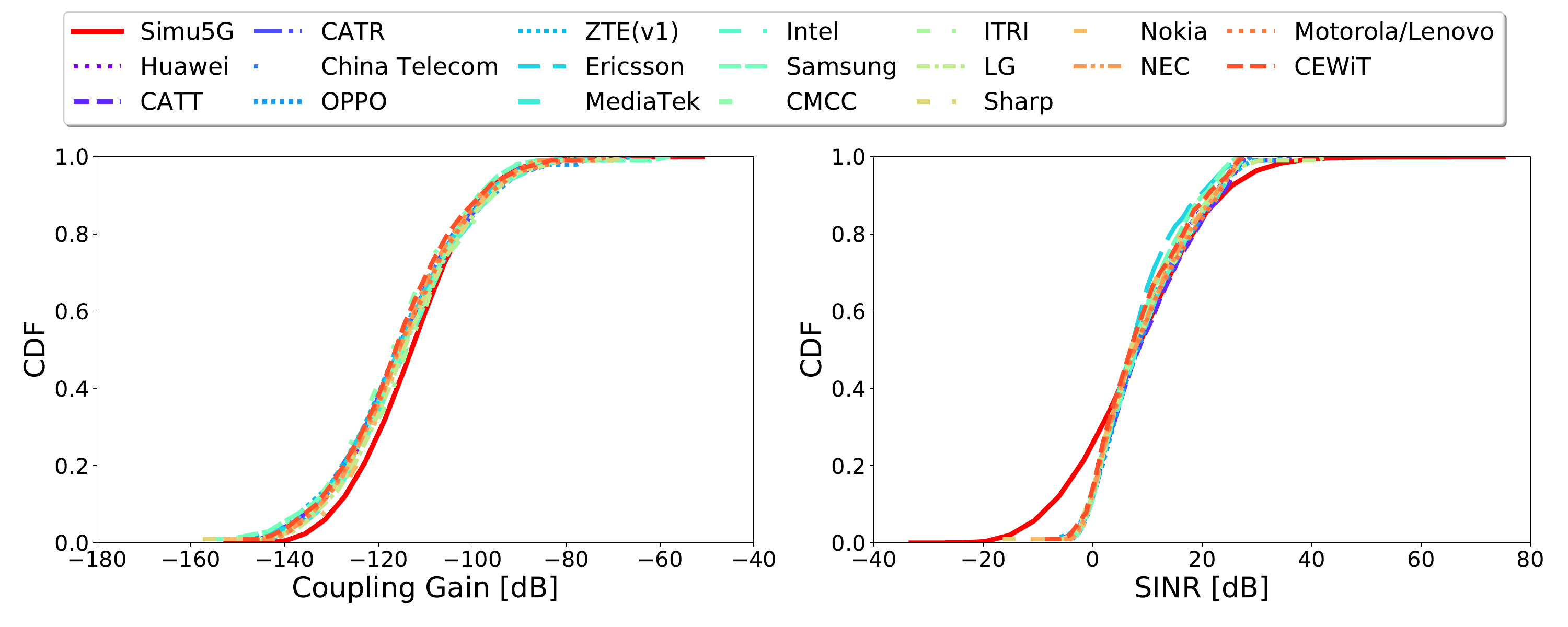}
\caption{Simu5G results compared to 3GPP submissions for the Rural eMBB.}
\label{fig:rural}
\end{figure*}

In rural deployment, both KPIs have a very good alignment with the 3GPP submissions, which is depicted in Fig.~\ref{fig:rural}. Specifically, we obtain slightly higher values for Coupling Gain than the submissions. Note that the beamforming gains do not have a significant influence on the coupling gain for the rural scenario due to the lack of obstacles. 
The Wideband SINR has a similar trend as in the urban deployment with a close match in the mid-range, while the two extremes deviate from the submissions. The lower range discrepancy may occur for the same reasons we observed in the urban deployment, i.e., the lack of beamforming still can lead to higher levels of interference between cells. In the case of the higher range, it may happen because Simu5G completely separates downlink and uplink when calculating the interference without connecting uplink users to the same base station \cite{nardini2020simu5g}. Thus, the downlink bands could be slightly less occupied to generate the obtained outliers. 
The KS scores (Table~\ref{tab:ks_rural}) also match the CDF trend, where SINR is relatively close to the average value. In the Coupling Gain, the values are very close to the maximum and average.

\begin{table}[!ht]
    \centering
    \large
    \resizebox{\columnwidth}{!}{%
    \begin{tabular}{|c|c|c|}
         \hline
        Name & Coupling Gain Score & Wideband SINR Score   \\
        \hline
        Simu5G & 0.089 & 0.160 \\
        Average Value in 3GPP & 0.034 & 0.053 \\
        Maximum Value in 3GPP & 0.069 & 0.366 \\
        \hline
    \end{tabular}}
    \caption{Kolmogorov–Smirnov test results for the Rural eMBB.}
    \label{tab:ks_rural}
\end{table}

%% file: sections/automation.tex
\section{Automating Configuration}

Simu5G users (\textit{SimuU}) need to mainly manipulate \textit{Initialization} and \textit{Network Description (NED)} files as in Omnet++, where the former is for initializing parameters (e.g., transmission power and frequency). On the contrary, the topological configuration (e.g., number and types of devices) is specified in the NED file. The configuration complexity of these files may require users to have in-depth knowledge of various parameters (e.g., number of bands, device heights, thermal noise levels, angle of transmissions, type of penetration losses) and their settings, which can be time-consuming and error-prone. Furthermore, Simu5G supports different deployment environments (e.g., rural or urban), which require different parameter settings, including the default ones. For instance, the base station height in the rural area is higher than that of the urban one.  

Thus, we developed an API to generate configuration files for Simu5G automatically. SimuU only needs to provide high-level parameters and topological information, which the API will then convert to corresponding Simu5G configurations. Note that this automation capability is appended to the calibrated version of Simu5G to comply with 3GPP.

\subsection{Automation Workflow}

\begin{figure} [!ht]
    \centering
    \includegraphics[width=0.45\textwidth]{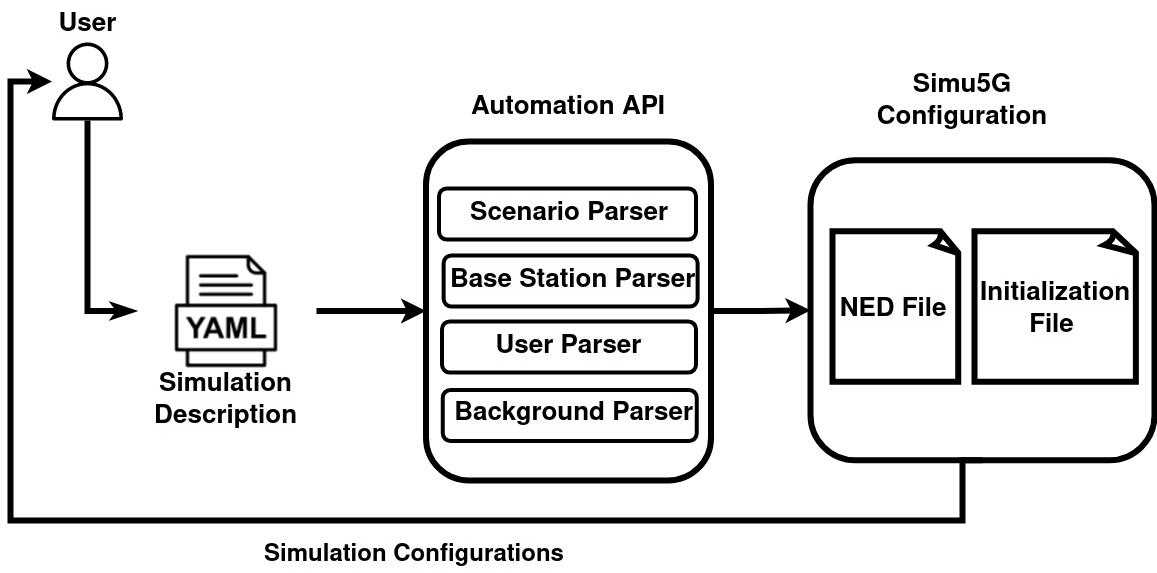}
    \caption{Workflow of the proposed automation API.}
    \label{fig:automation_workflow}
\end{figure}

Fig.~\ref{fig:automation_workflow} presents the proposed automation workflow, where users define their configuration demand using YAML \cite{yaml}. It is a widely used human-readable data serialization language in defining configuration files. Based on the provided YAML descriptions, we have developed a script to parse and generate the necessary Simu5G configurations. In the following, we present the key functionalities and their configuration using the proposed tool. Note that SimuU can extend the automation tool to add new configuration tasks. 


\textbf{Deployment scenario and simulation time.} These are the first set of parameters SimuUs need to define, where the deployment scenario can be urban or rural with options eMBB, mMTC, or URLLC.  We introduce \textit{scenario} and \textit{simulation-time} parameters to the defined scenario and time in the YAML file.  

\textbf{Base station positioning and sector configuration.} 
Next, SimuU must define the position of the chosen number of base stations and their respective number of sectors.

\textbf{Base station X2 connections.} Another configuration that can be quite extensive in the Simu5G configuration files is the X2 connections definitions. These links define the communication channels between base stations used substantially in processes like handover. In the Simu5G configuration, SimuUs need to define the number of ports available on each base station as well as define the link in the NED file. In the proposed tool, they simply need to provide the link numbers to enable connection among base stations. 


\textbf{User positioning.} In the case of user positioning, the proposed tool allows SimuU to define the number of users and the maximum allowed distances from the associated base stations. The tool then randomly deploys the specified number of users around each base station, respecting the defined max distance.

\textbf{Background user and base station positioning.} The SimuU can provide the number of background base stations and users to be used in Simu5G to simulate user interference. SimuUs simply need to provide the intended number of base stations and users along with the deployment area.

\textbf{KPIs collection.} In this case, SimuUs simply need to state the name of the KPIs that they want to monitor; thus, these are set in the configuration files to be collected. The defined KPIs are then stored in the standard ".vec" files, the default format in Simu5G.


The proposed automation tool uses four modules to parse the above inputs and generates the necessary configuration files for Simu5G, as shown in Fig.~\ref{fig:automation_workflow}.

The first module is the \textit{Scenario Parser}, which handles the KPI definitions, the simulation time, and deployment scenarios. Specifically, the tool turns on all mentioned KPIs by setting the corresponding flag to \textit{true}. In the case of scenarios, the tool sets the default parameters that 3GPP suggested along with the calibrated parameter values, as seen in Table \ref{tab:scenario}. We then set the simulation period as mentioned in the input file.  

The next module is the \textit{Base Station Parser}, which handles the base station positioning, sector configurations, and X2 connections. For the sector configuration, Simu5G does not have the option to define the number of sectors in a cell site; thus, the proposed tool defines multiple stations with different transmission angles. The parser first reads the configurations for the number of sectors and cell site positions, and creates the sites first and then the corresponding number of sectors with the defined angles. Specifically, the tool specifies the positions of the sites and sets the defined transmission angles for their sectors in the Simu5G initialization file following the 3GPP standard \cite{Huawei}. 

For the X2 connections, the module first reads all connection information. Then, it creates pairwise connections between stations by setting the NED file. After this, it calculates the number of necessary ports (one for each connecting station) for each base station and creates the port configuration in the Initialization file. Specifically, we start with the lowest available port number and keep connecting all required base stations.

The third module \textit{User Parser} first calculates the total number of users and sets the Initialization file accordingly. Then, for every base station, it calculates the deployment boundary of each user using the provided maximum distance in the YAML file. Based on that calculation, the initial position for the users is determined and set in the Initialization file.  


The fourth module is the \textit{Background Parser}, where all background base stations and user configurations are managed. This module parses the background base station positions based on the provided deployment boundaries. Specifically, their positions are randomly chosen within the given boundaries. Similarly, the tool parses the number of users, i.e., the provided number of background users in the Initialization file. Then, it sets their initial positions to be random ones within the defined ranges.


\subsection{Implementation and Results}

We implemented the automation tool using Python. It takes the YAML file as input and generates the Simu5G configuration files as output. We implemented both the urban-eMBB and rural-eMBB configurations calibrated in Section~\ref{sec:calibration}. The Simu5G initialization file for the urban-eMBB scenario normally would have over 1000 lines of code, which the proposed automation tool configures using only around 100 lines of the YAML input file, making setting up the experiments simpler and quicker. We share these configuration files in \cite{Boeira_Calibration-and-Automation-of-a-5G-Simulator_2024}. Due to space limitations, we present two examples of automation outcomes below to illustrate the benefits of such a tool.



\begin{figure} [!ht]
\centering
\begin{subfigure}[c]{0.3\columnwidth}
  \centering
  \includegraphics[width=0.6\linewidth]{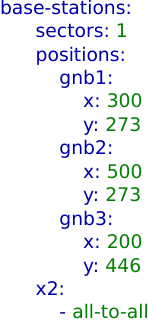}
  \caption{YAML code}
  \label{fig:yaml}
\end{subfigure}%
\begin{subfigure}[c]{0.65\columnwidth}
  \centering
  \includegraphics[width=1\linewidth]{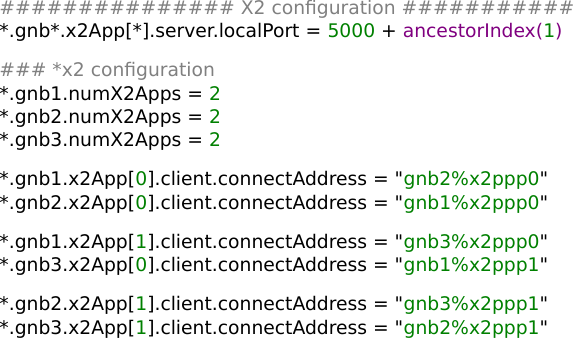}
  \caption{Simu5G configuration.}
  \label{fig:ini}
\end{subfigure}
\caption{An example of using Simu5G configuration automation to generate X2 connections.}
\label{fig:automation_x2}
\end{figure}

Fig.~\ref{fig:automation_x2} shows an example automation scenario, where SimuU defined the number, locations, and X2 connections of base stations. The corresponding X2 configuration in Simu5G configuration is shown in Fig~\ref{fig:automation_x2}(b). As you can see, users do not need to know the detailed structures of the simulators, which the proposed tool has taken care of. For instance, users simply need to define the X2 connection as "all-to-all" in the YAML file, and the tool does the rest. 


\begin{figure} [!ht]
\centering
\begin{subfigure}[c]{0.3\columnwidth}
  \centering
  \includegraphics[width=0.95\linewidth]{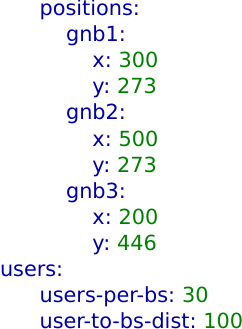}
  \caption{YAML code}
  \label{fig:yaml}
\end{subfigure}%
\begin{subfigure}[c]{0.65\columnwidth}
  \centering
  \includegraphics[width=0.95\linewidth]{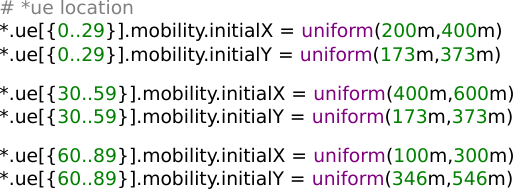}
  \caption{Simu5G configuration.}
  \label{fig:ini}
\end{subfigure}
\caption{An example of using Simu5G configuration automation to automatically configure user deployment.}
\label{fig:automation_user}
\end{figure}

Another example of usage of this tool can be seen in Fig.~\ref{fig:automation_user}, where a SimuU needs to define only the number of users per base station and the maximum distance from their assigned cell sites. With that, the automation tool calculates the \textit{x} and \textit{y} boundaries around the station site and uses these values to deploy the users.

%% file: sections/use-cases.tex
\section{A case study: ML-based anomaly detection}


This section presents a use case of the calibrated Simu5G simulator by developing a neural network-based anomaly detection system in 5G RAN. A RAN can suffer from faults or anomalies like suboptimal parameter configuration, interference, etc., and degrade the targeted service level objectives (SLOs). Thus, it is crucial for operators to detect and fix such anomalies on time, which can be done using a learning-based approach due to the complexity of the network and the availability of data. We can use historical KPI data (RSRP, RSRQ, SINR, etc.), to detect or predict anomalies in 5G RAN. In the following section, we present how the proposed calibrated and automated simulator can support data collection of KPIs for both, normal and faulty cases, to develop a learning-based anomaly prediction system.

\subsection{Dataset Description and Pre-processing} 

We generate time series KPIs for both normal and anomalous using the calibrated Simu5G. Specifically, we consider three common failures: excessive power reduction, too-late handover, and interference. The percentage of failures or anomalies is kept at most 2\% following our partner's recommendation, which also aligns with other industrial recommendations. We label the failed data while generating the KPIs. Our data set consists of around three million KPI data points, which include UE positions, serving cell distance, reference signal received power (RSRP), reference signal received quality (RSRQ), signal-to-interference-plus-noise ratio (SINR). 

We aggregate user-level time series KPIs to base station-level time series KPIs:
1) The total simulation time is binned into one-second time intervals. We perform average aggregation over time intervals to get time series KPI data for each time interval for the UEs.
2) At each time interval, we take the average aggregation of KPI data across all UEs a base station serves. This gives the time series KPI data for discrete time intervals for all base station sectors in the network.
3) We average across sectors, which gives the time series base station KPI data.

\subsection{Model Construction}
Historical KPI data is used for anomaly detection in 5G RAN base stations. So, we need models that can capture both the inter-base station spatial dependencies and per base station temporal patterns. To meet these requirements, we implement state-of-the-art Graph Convolution Network (GCN) \cite{kipf2016semi} and Multivariate Time Series Graph Neural Network  \cite{wu2020connecting} to detect anomalies in 5G RAN over the collected KPIs. We compare the two models and go over their architectural details. 

\textbf{GCN.} Graph Convolution Network \cite{kipf2016semi} is tailored for processing graph-structured data, where the graph convolutional layer is adapted to work on irregular graph structures. Unlike Convolutional Neural Network (CNN) operating on regular grids, GCNs aggregate information from neighboring nodes through message passing. The embedding process, transforming high dimensional data into meaningful low dimensional vectors \cite{zhang2019heterogeneous}, involves encoding node features using their neighborhood relationships. We use two graph convolution layers to learn embeddings, Relu activation function to introduce non-linearity, and fully connected layers to map learned features to the output space. 
Thus, GCN models node representations of base stations, to perform anomaly prediction based on neighborhood. Even though this method captures important inter-base station patterns, it does not capture temporal patterns from historical KPI data. There are no modules in the GCN model for time series modeling; it can only rely on the base stations' neighborhood to make predictions, which is not ideal. This issue is mitigated in MTGNN.

\textbf{MTGNN.} We also implement MTGNN \cite{wu2020connecting}, an architecture designed for multivariate time series data in graph data, for anomaly detection in 5G RAN. There are three primary components of the MTGNN architecture: graph learning, graph convolution, and temporal convolution. 1) The graph learning module generates an adjacency matrix adaptively to capture hidden relationships in time series data; it can work with or without explicit graph structure. In our use case, we do not know any information about the interdependencies among base stations. So, there is no structure provided while using MTGNN. 2) The graph convolution module collects nodes' neighborhood information by utilizing a two step mix hop propagation layer that involves information propagation along the graph structure and information selection at each hop. 3) The temporal convolution module captures time series patterns from historical KPI through the use of dilated 1D convolution filters. This allows the module to discover patterns at different ranges and handle long sequences by utilizing multiple filter sizes and dilated convolution. With these three modules MTGNN models both base station neighborhood influence and historical time series KPI data for anomaly prediction. This gives MTGNN a clear advantage over GCN, which is evident in the experimental results discussed in the following section.

\subsection{Results}

We report the precision, recall, and F1-score averaged across the classes on the test dataset, which is generated using the calibrated Simu5G. We train GCN and MTGNN and report the 5-fold cross-validation results on the test set. The F1-scores with corresponding precision and recall for the anomaly detection are shown in Table \ref{tab:result}. All the results demonstrate the implemented models' ability to predict anomalies successfully. We can see MTGNN performs noticeably better than GCN. This is likely because GCN has no component for modeling time series data. It only looks at the current time step KPI data to detect anomalies, while MTGNN uses convolution layers to capture temporal dependencies. This demonstrates a use case for reliably using calibrated Simu5G data for anomaly prediction.

\begin{table}
\centering
\caption{The performance of anomaly detection.}
\label{tab:result}
\begin{tabular}{|m{1.5cm}|m{1.5cm}|m{1.5cm}|m{1.5cm}|}
 \hline
 Models & Precision & Recall & F1-score \\ 
 \hline
 GCN & 0.99 & 0.62 & 0.62 \\ 
 \hline
 MTGNN & 0.99 & 0.75 & 0.74 \\ 
 \hline
\end{tabular}
\end{table}


%% file: sections/conclusion.tex
\section{Conclusion}

In this work, we have calibrated a system-level simulator, Simu5G, in strict adherence to the 3GPP specifications. The results have shown that the calibration outcome closely matches the 3GPP submissions.  Furthermore, we introduced an Application Programming Interface (API) designed to expedite the simulation process in Simu5G. This tool facilitates the swift configuration with standard parameters, providing users with a higher level of abstraction. This not only diminishes the learning curve associated with Simu5G but also enhances the efficiency of the simulation configuration and usage process. Finally, we present a use-case of the calibrated Simu5G by developing a machine learning-based anomaly prediction system using the data generated in the calibrated and automated Simu5G.

In future works, we intend to expand our calibration scenarios to feature other configurations for urban and rural scenarios, such as configurations B and C. Additionally, we aspire to extend our calibration efforts to encompass various scenarios, including the Indoor Hotspot scenario as well as scenarios related to Ultra-Reliable Low Latency Communications (URLLC) and Massive Machine Type Communications (mMTC). Besides that, we want to utilize the data-generation capabilities of the simulator to expand on our machine learning-based models and perform Root-Cause Analysis and automatic repair in these scenarios.
